\documentclass{article}
\usepackage{PRIMEarxiv}

\usepackage{amsmath}
\usepackage{amssymb}
\usepackage{mathtools}
\usepackage{amsthm}
\usepackage{lineno,hyperref}
\usepackage{times}
\usepackage{soul}
\usepackage{url}
\usepackage[small]{caption}
\usepackage{graphicx}
\usepackage{amsthm}
\usepackage{booktabs}
\usepackage{amsfonts}
\usepackage{algorithmicx}
\usepackage{algorithm}
\usepackage[noend]{algpseudocode}

\theoremstyle{plain}

\theoremstyle{definition}

\theoremstyle{remark}

\title{Multi-Agent Cooperation via Unsupervised Learning of Joint Intentions}

\author{%
  Shanqi Liu\\
  Control Science and Engineering \\
  Zhejiang University, Zhejiang, China\\
  \And
  Weiwei Liu \\
  Control Science and Engineering \\
  Zhejiang University, Zhejiang, China\\
  \And
  Wenzhou Chen \\
  Computer Science \\
  Hangzhou Dianzi University, Zhejiang, China\\
  \And
  Guanzhong Tian \\
  Ningbo Research Institute\\
  Zhejiang University, Zhejiang, China\\
  \And
  Yong Liu \\
  Control Science and Engineering \\
  Zhejiang University, Zhejiang, China\\
}

\begin{document}

\maketitle

	\begin{abstract}
		The field of cooperative multi-agent reinforcement learning (MARL) has seen widespread use in addressing complex coordination tasks. While value decomposition methods in MARL have been popular, they have limitations in solving tasks with non-monotonic returns, restricting their general application. Our work highlights the significance of joint intentions in cooperation, which can overcome non-monotonic problems and increase the interpretability of the learning process. 
		To this end, we present a novel MARL method that leverages learnable joint intentions. Our method employs a hierarchical framework consisting of a joint intention policy and a behavior policy to formulate the optimal cooperative policy.  The joint intentions are autonomously learned in a latent space through unsupervised learning and enable the method adaptable to different agent configurations. Our results demonstrate significant performance improvements in both the StarCraft micromanagement benchmark and challenging MAgent domains, showcasing the effectiveness of our method in learning meaningful joint intentions. 
	\end{abstract}

\section{Introduction}
In recent years, cooperative Multi-Agent Reinforcement Learning (MARL) has gained significant attention as a prominent approach for learning effective behaviors in real-world tasks, such as swarming \cite{huttenrauch2017guided} and autonomous driving \cite{cao2012overview}. These tasks involve multiple agents working in collaboration within a shared environment.
The most prevalent framework for MARL is Centralized Training with Decentralized Execution (CTDE) methods \cite{jiang2018graph}, which addresses practical communication constraints and handles the exponentially growing joint action space. However, CTDE methods still face challenges in real-world applications.
In particular, the popular value-based CTDE algorithm, QMIX \cite{rashid2018qmix}, may not always approximate the optimal value function, as the monotonicity constraint leads to sub-optimal value approximations, also referred to as the relative overgeneralization problem \cite{panait2006biasing}. This results in QMIX being unable to capture value functions that are dependent on other agents' actions. Additionally, current methods tend to perform poorly in ad-hoc team play scenarios \cite{stone2010ad}, where the environment has varying team sizes at test time. The reason for this is that agents must assess and adapt to others' capabilities in order to exhibit optimal behavior in the ad-hoc team play MARL setting \cite{zhang2020multi}. However, current methods tend to learn fixed policies, which is not suitable for these dynamic scenarios.

A potential solution to the relative overgeneralization issue is to incorporate joint intentions into the MARL framework \cite{jennings1995controlling}. In human cooperation, individuals often work towards a common goal without direct access to each other's actions by aligning with a shared joint intention. In MARL, if agents can coordinate their actions based on a shared joint intention, they can choose optimal cooperative actions without requiring information about each other's specific actions, as they are working towards the same purpose. For example, in a \textit{pursuit} task where multiple agents must work together to attack a prey and a single attack incurs a penalty, if agents can coordinate their actions based on a joint intention such as all attacking together, they can avoid a situation where one agent is afraid of another not attacking, leading to a decision to stay to avoid incurring the penalty. Nevertheless, defining joint intentions is a challenge and manually specified joint intentions lack generalizability. Hence, this work proposes a MARL approach that utilizes unsupervised learning to derive joint intentions.

Our approach for achieving unsupervised learning of joint intentions in MARL consists of the following steps:
\begin{itemize}
\item Team partitioning: We partition the team of agents into smaller groups and share joint intentions within each team.

\item Hierarchical framework: The approach consists of two levels, a high-level joint intention policy and a low-level behavior policy. The joint intention policy chooses the optimal joint intention from a latent space based on the local state of each team. The behavior policy, on the other hand, is used by every agent in the team to take the optimal action corresponding to the team's joint intention.

\item Unsupervised learning of joint intentions: We optimize the mutual information between the joint intentions shared within each team and the changes in the local state of each team. This allows us to map different joint intentions to specific changes in the local state.

\item Adaptation to dynamic agent numbers: The learned joint intentions are independent of the number of agents in the team, allowing the approach to adapt to environments with uncertain agent numbers.

\item Overcoming non-stationarity: We extend the approach by utilizing mutual information to increase the expressiveness of the mixing network of the joint intention policy's value function, thereby overcoming the non-stationarity of the hierarchical framework.
\end{itemize}
Experimental results show that our approach outperforms state-of-the-art methods in both monotonic and non-monotonic scenarios in games such as StarCraft and MAgent, as well as in ad hoc team play environment settings. The visualization of joint intentions representations demonstrate the meaningfulness of the learned joint intentions and their impact on agent behavior. Additionally, the analysis of the evolution process of the learned joint intentions highlights their relevance to performance improvement and provides interpretability for the cooperation among agents.

\section{Related Work}
Currently, the mainstream methods are CTDE \cite{lowe2017multi,iqbal2019actor} methods in MARL.
In CTDE methods, VDN \cite{sunehag2017value} learns the joint-action Q-values by factoring them as the sum of each agent's utilities. QMIX \cite{rashid2018qmix} extends VDN to allow the joint action Q-values to be a monotonic combination of each agent's utilities that can vary depending on the state. 
However, the monotonic constraints on the joint action-values introduced by QMIX and similar QMIX-class methods lead to provably poor exploration and relative overgeneralization \cite{panait2006biasing}. 
To address this problem, QPLEX and QTRAN \cite{wang2020qplex, son2019qtran} aim to learn value functions with complete expressiveness capacity. However, they are reported to perform poorly when used in practice because learning complete expressiveness is impractical in complicated MARL tasks due to the difficult exploration in large joint action space \cite{gupta2021uneven,wan2021greedy}. 
Other methods involve additional structure to facilitate the convergence of the value function to optimum, such as MAVEN \cite{maven} hybridizes value and policy-based methods by introducing a latent space for hierarchical control. This allows MAVEN to achieve committed, temporally extended exploration. 
UneVEn \cite{gupta2021uneven} learns a set of related tasks simultaneously with a linear decomposition of universal successor features to generalize to the non-monotonic tasks. 
Moreover, G2ANET \cite{liu2020multi} models the importance of relationship between agents by a complete graph and a two-stage attention network. However, these methods lack optimality guarantees and are insufficient to thoroughly address the relative overgeneralization problem \cite{wan2021greedy}.
Additionally, all these methods can not estimate the value of actions considering the changing of other agents' actions, which is essential for cooperation in environments with non-monotonic returns. Our method learns the joint intentions shared among agents, enabling agents to cooperate based on awareness of other agents' intentions to completely avoid the mis-coordination that leads to the relative overgeneralization. 
Moreover, our method can adapt to ad hoc settings and the learned joint intentions provide interpretability that other methods lack. 
Additionally, compared to other methods that learn to model others' intentions or behaviors like \cite{strouse2018learning,kim2020communication}, our method uses the joint intention as a constraint to avoid mis-coordination instead of predicting other agents' behaviors. 

Our work is also related to information theory. 
Related work SAC-AWMP \cite{hou2020off} has verified the effectiveness of information theory in the RL training process. 
DIAYN \cite{eysenbach2018diversity}, DADS \cite{sharma2019dynamics} use mutual information as an intrinsic reward to discover skills without an extinct reward function. 
To optimize the mutual information, \cite{hausman2018learning} showed that a discriminability objective is equivalent to maximizing the mutual information between the desired objectives. 
In MARL, ROMA \cite{wang2020roma} use mutual information to learn roles for agents corresponding to the past trajectories. 
\cite{jaques2019social} learns social influence between agents by estimating the mutual information between their actions. 
Similarly, \cite{wang2019influence} uses an intrinsic reward to characterize and quantify the influence of one agent’s behavior on the expected returns of other agents. 
However, these methods do not aim to address the relative overgeneralization problem. Our method differs from all these methods. We propose unsupervised learning of joint intentions shared within teams and learn the joint intention-based multi-agent policies to address the relative overgeneralization problem. 

\section{Background}
A fully cooperative multi-agent sequential decision-making task can be described as a decentralized partially observable Markov decision process (Dec-POMDP), which is defined by a set of states $S$ describing the possible configurations of all N agents, a set of possible actions $A_1, . . . , A_N$, and a set of possible observations $O_1, . . . , O_N$. At each time step, each agent $i \in \{1, ..., N\}$ chooses an action $a_i \in A_i$, forming a joint action $\mathbf{u} \in U$. The joint action $\mathbf{u}$ produces the next state by a transition function $P_s:S \times U \rightarrow S$. The next observation of each agent $o \in O$ is updated by an observation function $O_p:S \rightarrow O$. All agents share the same reward $r:S \times U \rightarrow R$ and with a joint value function $Q_{tot} = E_{s_{t+1}: \infty ,a_{t+1}: \infty } [R_t|s_t, \mathbf{u_t}]$ where $ R_t = \sum^{\infty}_{j=0} \gamma^j r_{t+j}$ is the discounted return.
Partial observability is typically handled by using the history of actions and observations as a proxy for state, often processed by a recurrent neural network: $Q_{tot}(\mathbf{\tau_t}, \mathbf{u_t}) \approx Q_{tot}(s_t, \mathbf{u_t})$, where $\tau_i^t$ is $(o_i^0, a_i^0, ..., o_i^t)$ and $\mathbf{\tau_t} = \{\tau_i^t\}_{i \in A}$.
In this work, we also introduce a set of latent joint intentions $Z$ describing diverse modes of changes of local state, agents in the same team $j$ share the same $z_j \in Z$. And the joint intentions can be viewed as $\mathbf{z}$. We use $N,M,K_j$ to refer to the number of agents, the number of teams and the number of agents in team $j$. 

Moreover, non-stationarity in hierarchical reinforcement learning means the transition collected when the low level policy is not well-trained could be useless for high level policy because the low level policy is constantly changing and the collected transition could be different under the current low level policy. 

\section{Method}

\begin{figure*}[t]
	\centering
	\includegraphics[width=0.95\linewidth]{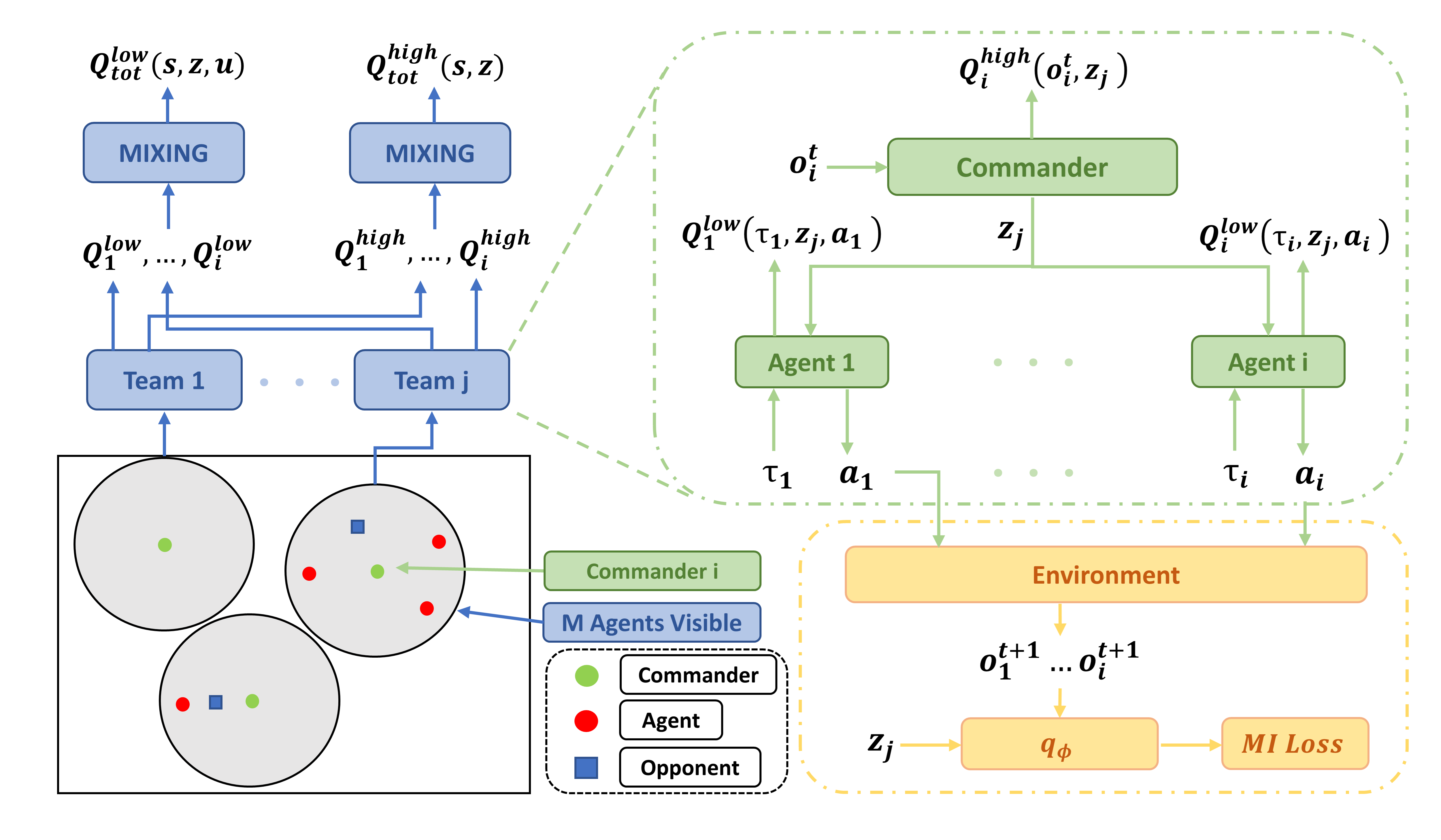}
	\caption{The architecture of our method. Left: The architecture for total Q-values and team partition. Upper Right: The hierarchical framework of each team. Lower Right: The optimization of mutual information objective.}
	\label{method}
\end{figure*}

\subsection{Team Partitioning}
Our method is designed to build a multi-agent framework that can make the agents in the same team share the same joint intention to cooperate better. Since we know the joint intentions of humans are usually related to describing the environment changes, like passing the ball to the center forward, which means the center striker and center forward work together to change the ball's position. 
Therefore, the joint intention shared within the team should also describe the changes of the environment. Furthermore, since our joint intentions are shared within teams, instead of describing the changes of the global state, our joint intention describes the changes of the local state of the team itself. 
However, it is difficult to determine how to divide all agents into distinct teams to better present joint intentions. 
To address this problem, we propose a team partitioning approach from the perspective of reward decomposition. 

\textbf{Theorem 1:} 
\textit{For all $i \in \{1,...,N\}$, there exists a reward decomposition $\hat{r_i}$ that only depends on $(o_i^t, u^{t-}_i, a_i^t)$ where $u^{t-}_i$ is the joint actions of agents who can be observed by agent $i$, so that
\begin{equation}
    \begin{array}{c}
        r_{tot}(s_t, \mathbf{u_t}) = \sum_{i=1}^{N} (\hat{r_i}(o_i^t, u^{t-}_i, a_i^t) + (2^{(N-K_i)}-1) \times \epsilon).
    \end{array}
\end{equation}
where $r_{tot}$ is the external reward of the environment, $K_i$ is the number of agents that can be observed by agent $i$, and $(2^{(N-K_i)}-1) \times \epsilon$ is the optimality gap where the $\epsilon$ is a small value. 
}

\textbf{Proof.}

First of all, as the total reward of the environment are generated by all kind of possible interaction between agents, we have
\begin{equation}
    \begin{array}{c}
        r_{tot}(s_t, u_t) = \sum_{i=1}^{N} r_1(o_i^t,a_i^t) + \sum_{i<j=2}^{i<j=N} r_{2}(o_i^t, o_j^t, a_i^t, a_j^t) + ... + r_{N}(o_1^t,..., o_N^t, a_1^t,..., a_N^t).
    \end{array}
    \label{eq1a}
\end{equation}
Notable, each item $r_k$ in Eq. (\ref{eq1a}) exists only when there are interactions between the described agents. Otherwise, it should equal to zero as the reward is represented by other items.
Especially, in decentralized execution settings, we assume that the credit assignment can only learn the reward decomposition that each agent interacts with the agents who can be observed by themselves. The reason is that the decentralized policy can only take action according to information within the observation. If the Q-value function based on the learned reward decomposition needs extra information to be calculated, we must introduce the communication method to transfer the necessary messages. Therefore, for any  $r_k$ in Eq. (\ref{eq1a}) that represents the reward of agents which are not in the view of each other, we model the item as an unlearnable noisy $W(x)$ with the expectation $\epsilon$. 
Additionally, most scenarios have a smaller range of interaction than the field of view in practice, which means that the possibility of agents having interaction with agents out of view is low. Therefore, the $\epsilon$ is a small value. 



Next, we define $\hat{r_i}(o_i^t, u_i^{t-}, a_i^t)$ and decompose it similarly as,
\vspace{-5pt}
\begin{equation}
    \begin{array}{c}
        \hat{r_i}(o_i^t, u_i^{t-}, a_i^t) = r_1^i(o_i^t, a_i^t) + \sum_{j=1,j \neq i}^{K_i} r_2^i(o_i^t, a_j^t, a_i^t)
         + ... + r_{K_i}^i(o_i^t, a_1^t,..., a_{K_i}^t, a_i^t).
    \end{array}
    \label{eq3a}
\end{equation}
We notice that for $r_k^i = r_k^i(o_i^t, a_j^t,...a_k^t, a_i^t)$ if $r_k^i$ is not zero, we have 
\begin{equation}
    \begin{array}{c}
        r_k^i(o_i^t, a_j^t,...a_k^t, a_i^t) = \frac{1}{k} r_k(o_i^t, o_j^t,...,o_k^t, a_j^t,...a_k^t, a_i^t).
    \end{array}
    \label{eq4a}
\end{equation}
The reason is that if $r_k$ exists, it represents the cooperation reached by agents set $(i,j, .., k; j<i<k)$. Moreover, as the specific joint action is taken by all agents in the set, each agent should equally share the cooperative reward $r_k$. Then, we can modify Eq. (\ref{eq3a}) as
\begin{equation}
    \begin{array}{c}
        \hat{r_i}(o_i^t, u_i^{t-}, a_i^t) = r_1(o_i^t, a_i^t) +  \frac{1}{2} \sum_{j=1,j  \neq i}^{K_i} r_2(o_i^t,o_j^t, a_j^t, a_i^t)
         + ...  + \frac{1}{K_i} r_{K_i}(o_i^t,o_1^t,...,o_{K_i}^t, a_1^t,..., a_{K_i}^t, a_i^t).
    \end{array}
    \label{eq5a}
\end{equation}
We construct the following equation,
\begin{equation}
    \begin{array}{c}
        \sum_{i=1}^{N} (\hat{r_i}(o_i^t, u^{t-}_i, a_i^t) + (2^{(N-K_i)}-1) \times \epsilon) \\= \sum_{i=1}^{N} r_1(o_i^t, a_i^t) +  \sum_{i=1}^{N} \frac{1}{2} \sum_{j=1,j  \neq i}^{K_i} r_2(o_i^t,o_j^t, a_j^t, a_i^t)
         + ... + \sum_{i=1}^{N} (C_{(N-K_i)}^{2} + C_{(N-K_i)}^{3} +... + C_{(N-K_i)}^{N-K_i}) \times \epsilon
          \\=\sum_{i=1}^{N} r_1(o_i^t, a_i^t)+ \frac{1}{2} \sum_{i=1}^{N} \sum_{j=1,j  \neq i}^{K_i} r_2(o_i^t,o_j^t, a_j^t, a_i^t) + \sum_{i=1}^{N} \sum_{j=1,j  \neq i}^{N-K_i} \epsilon
          + ... 
    \end{array}
    \label{eqna}
\end{equation}
where we use the noisy function's expectation $\epsilon$ stands for the corresponding reward items which are related to agents out of view of agent $i$. These items can be considered as optimality gaps. 
Then, we have Eq. (\ref{eqna}) equals 
\begin{equation}
    \begin{array}{c}
        \sum_{i=1}^{N} r_1(o_i^t, a_i^t)+ \frac{1}{2} \sum_{i=1}^{N} \sum_{j=1,j  \neq i}^{K_i} r_2(o_i^t,o_j^t, a_j^t, a_i^t) + \sum_{i=1}^{N} \sum_{j=1,j  \neq i}^{N-K_i} \epsilon
        + ... 
          \\ = \sum_{i=1}^{N} r_1(o_i^t, a_i^t)+ \frac{1}{2} \sum_{i=1}^{N} \sum_{j=1,j  \neq i}^{N} r_2(o_i^t,o_j^t, a_j^t, a_i^t) + ... 
        \\ = \sum_{i=1}^{N} r_1(o_i^t, a_i^t)+ \frac{1}{2} \sum_{i<j=2}^{i<j=N} 2 \times r_2(o_i^t,o_j^t, a_j^t, a_i^t) + ... 
        \\ = \sum_{i=1}^{N} r_1(o_i^t, a_i^t)+ \sum_{i<j=2}^{i<j=N} r_2(o_i^t,o_j^t, a_j^t, a_i^t) + ... = r_{tot}(s_t, u_t).
    \end{array}
\end{equation}
So, we have 
\begin{equation}
    \begin{array}{c}
        r_{tot}(s_t, u_t) = \sum_{i=1}^{N} (\hat{r_i}(o_i^t, u^{t-}_i, a_i^t) + (2^{(N-K_i)}-1) \times \epsilon).
    \end{array}
    \label{eq6a}
\end{equation}
Therefore, we have proved our Theorem 1. \qed

Since $\hat{r_i}$ only depends on agent $i$ and agents in the view of agent $i$, we notice that if we use the joint intention to describe the change of the local observation of agent $i$, then the joint intention is only related to agent $i$ and agents in the view of agent $i$. This result indicates that we can use a commander-follower structure to partition agents that the commander agent $i$ uses its observation as the local state of the team and the team members can be determined as all agents in the view field of the commander. 

\begin{algorithm}[t]
	\caption{Greedy iteration of team partitioning}




	\begin{algorithmic}[1]

			\For{each time step}

			Initialize to put all agents into an unassigned set $\mathbf{S}$

			\While{S}

			Find the agent $\mathbf{A_i}$ with most agents $\mathbf{K_i}$ in view field 

			$\mathbf{T_i = A_i + K_i}$ where $\mathbf{T_i}$ means team $i$

			$\mathbf{S = S - T_i}$

			\EndWhile

			\EndFor
	\end{algorithmic}
	\label{algo}
\end{algorithm}

However, it is still not clear how to choose the commander. From Theorem 1, we can see that the optimality gap heavily depends on the number of observed agents. If the local observation of agent $i$ covers more agents, then the gap is smaller. Following this principle, we choose the agents which minimize the optimality gap as commanders. However, the searching time of finding the optimal solution is exponential, so in practice, we use a greedy algorithm to find a partition. 
We first add all agents into an unassigned set during each time step. Then, we always find the agent who can observe the most agents as a team commander and make all agents who are being observed as the team members. Then, we remove all these agents from the unassigned set and repeat the iteration approach until no agents are left. Notably, we randomly choose the commander when multiple agents can observe the same number of agents. The algorithm is included in Algorithm \ref{algo}. 
This approach can guarantee that every agent belongs to a specific team at each time step and the overall optimality gap is nearly minimized.
Intuitively, the team partition method chooses agents who can observe the most agents as the commanders. This is similar to the natural way of choosing commander that humans usually choose people with the most information about team members as the commander \cite{prentice2004understanding}. 
More intuitive representations of the team partition results are shown in Fig. \ref{method}. 

\subsection{Unsupervised Learning of Joint Intentions}
Since the team partitioning is clear, we now discuss the joint intention shared within the team to address the relative overgeneralization problem. 
As the joint intention makes all agents in the team working towards the same goal. This requires joint intention to have four properties:
\begin{itemize}
    \item [1)] Executable: All agents in the team must understand the joint intention and take the corresponding actions.
    \item [2)] Dynamic: The joint intention should be determined based on the states of each time step's teams.
    \item [3)] Diverse: Each joint intention is expected to describe one specific change of the local state.
    \item [4)] Learned: The joint intention is expected to be unsupervised autonomously learned from a latent space, not relying on pre-defined knowledge.
\end{itemize}
To implement these properties, we propose a hierarchical learning framework and an unsupervised learning approach. 
\subsubsection{Hierarchical Learning Framework}
The hierarchical policy of each agent consists of a high level joint intention policy and a low level behavior policy. The high level joint intention policy is designed to choose a joint intention $z$ from a latent space to maximize the return of the environment. The latent space is designed to map to diverse modes of local state changes, which are different joint intentions. We consider a discrete latent joint intention space as the size of the latent joint intention space should be small, otherwise it will lose generality. Therefore, for a team $j$ where agent $i$ is the commander, we model the joint intention policy as $\pi_h(z_j|o_i^t;\theta)$ parameterized by $\theta$, where $o_i^t$ is the observation of agent $i$. We sample $z_j \sim \pi_h(z_j|o_i^t;\theta)$ at each time step for each team during rollout. 

As for the low level behavior policy, it is expected to follow the joint intention given by the joint intention policy to take optimal cooperative actions. We model behavior policy as a joint intention conditioned policy $\pi_l(a_k^t|\tau_k^t,z_j;\mu)$ parameterized by $\mu$. Every agent in the team $j$ will follow the joint intention $z_j$ and take action $a_k$ which interacts with the environment. 
The joint actions $u_j$ of team $j$ can be viewed as $u_j \sim (\pi_l(a_0^t|\tau_0^t,z_j;\mu)...\pi_l(a_{K_j}^t|\tau_{K_j}^t,z_j;\mu))$, where $K_j$ means there are $K_j$ agents in team $j$. The overall hierarchical framework is shown in Fig. \ref{method}.

\subsubsection{Objectives of Optimization}
Introducing latent joint intention space and joint intention conditioned policy does not autonomously learn these four desired properties. To address this problem, we use the information theoretic paradigm of mutual information to learn joint intentions. Specifically, we propose to make behavior policy maximize the mutual information between the next local state $o_i^{t+1}$ and current joint intention $z_j$ conditioned on the current local state $o_i^{t}$.
\begin{equation}
	\begin{array}{c}
		I(o_i^{t+1} ; z_j \mid o_i^{t}) =H(z_j \mid o_i^{t})-H(z_j \mid o_i^{t+1}, o_i^{t}) 
		=H(o_i^{t+1} \mid o_i^{t})-H(o_i^{t+1} \mid o_i^{t}, z_j) = I(z_j ; o_i^{t+1} \mid o_i^{t}).
	\end{array}
\end{equation}
where $I$ is mutual information and $H$ is entropy. 
The meaning of such mutual information is that how much $z_j$ is related to the transition from $o_i^{t}$ to $o_i^{t+1}$. 
Therefore, having high mutual information means behavior policy can comprehend each joint intention's meaning and respond accordingly to make the transition described by the joint intention happen.
In other words, optimizing this mutual information can map different joint intentions in the latent space into specific changes of the local state. In this way, we can obtain the unsupervised autonomously learned joint intentions.
However, estimating and maximizing mutual information is often infeasible \cite{maven}. To address this problem, we introduce a variational posterior estimator which provides a lower bound for the mutual information \cite{alemi2016deep}. 
\begin{equation}
	\begin{array}{c}
		I(z_j ; o_i^{t+1} \mid o_i^{t})
		=\mathbb{E}_{z_j, o_i^{t}, o_i^{t+1}}[\log \frac{q_{\phi}(z_j \mid o_i^{t}, o_i^{t+1})}{p(z_j \mid o_i^{t})}] +\mathbb{E}_{o_i^{t}, z_j}[D_{K L}(p(z_j \mid o_i^{t}, o_i^{t+1})) \| q_{\phi}(z_j \mid o_i^{t}, o_i^{t+1})))] \\
		\geq \mathbb{E}_{z_j, o_i^{t}, o_i^{t+1}}[\log \frac{q_{\phi}(z_j \mid o_i^{t}, o_i^{t+1})}{p(z_j \mid o_i^{t})}].
	\end{array}
\end{equation}
where $q_{\phi}(z_j \mid o_i^{t}, o_i^{t+1})$ is the posterior estimator, parameterized by $\phi$ and $p$ is the true posterior. This lower bound can be estimated by optimizing the following loss of each team:
\begin{equation}
	\mathcal{L}_{I}^j= \mathbb{E}_{o_{i}^{t},o_{i}^{t+1}}\left[D_{\mathrm{KL}}\left[p\left(z_j \mid o_{i}^{t}\right) \| q_{\phi}\left(z_j \mid o_{i}^{t+1}, o_{i}^{t}\right)\right]\right].
\end{equation}
where $D_{\mathrm{KL}}$ is the KL divergence operator. 

\textbf{Proof.}

We will derive the posterior estimator in detail and find a tractable lower bound of the mutual information here.
Firstly, we have
\begin{equation}
	\begin{array}{c}
		I(z_j ; o_i^{t+1} \mid o_i^{t})
		=\mathbb{E}_{z_j, o_i^{t}, o_i^{t+1}}[\log \frac{p_{\phi}(z_j \mid o_i^{t}, o_i^{t+1})}{p(z_j \mid o_i^{t})}].
	\end{array}
\end{equation}
And we have
\begin{equation}
	\begin{array}{c}
		\mathbb{E}_{z_j, o_i^{t}, o_i^{t+1}}[\log \frac{p_{\phi}(z_j \mid o_i^{t}, o_i^{t+1})}{p(z_j \mid o_i^{t})}]
		=\mathbb{E}_{z_j, o_i^{t}, o_i^{t+1}}[\log \frac{q_{\phi}(z_j \mid o_i^{t}, o_i^{t+1})}{p(z_j \mid o_i^{t})}] +\mathbb{E}_{o_i^{t}, z_j}[D_{K L}(p(z_j \mid o_i^{t}, o_i^{t+1})) \| q_{\phi}(z_j \mid o_i^{t}, o_i^{t+1})))]
		\\ \geq \mathbb{E}_{z_j, o_i^{t}, o_i^{t+1}}[\log \frac{q_{\phi}(z_j \mid o_i^{t}, o_i^{t+1})}{p(z_j \mid o_i^{t})}].
	\end{array}
\end{equation}
For the last item, we have
\begin{equation}
	\begin{array}{c}
		\mathbb{E}_{z_j, o_i^{t}, o_i^{t+1}}[\log \frac{q_{\phi}(z_j \mid o_i^{t}, o_i^{t+1})}{p(z_j \mid o_i^{t})}]
		=\mathbb{E}_{z_j, o_i^{t}, o_i^{t+1}}\left[\log q_{\phi}\left(z_j \mid o_i^{t+1}, o_{i}^{t}\right)\right] +\mathbb{E}_{o_{i}^{t}}\left[H\left(z_j \mid o_{i}^{t}\right)\right] \\=
		\mathbb{E}_{o_i^{t+1}, o_{i}^{t}}\left[\int p\left(z_j \mid o_i^{t+1}, o_{i}^{t}\right) \log q_{\phi}\left(z_j \mid o_i^{t+1}, o_{i}^{t}\right) d z_j\right] +\mathbb{E}_{o_{i}^{t}}\left[H\left(z_j \mid o_{i}^{t}\right)\right]
		\\ =
		-\mathbb{E}_{o_i^{t+1}, o_{i}^{t}}[CE\left[p\left(z_j \mid o_{i}^{t}\right) \| q_{\phi}\left(z_j \mid o_i^{t+1}, o_{i}^{t}\right)\right]
		+ \mathbb{E}_{o_{i}^{t}}\left[H\left(z_j \mid o_{i}^{t}\right)\right].
	\end{array}
\end{equation}
where $CE$ means the cross entropy. In this way, we have
\begin{equation}
	\begin{array}{c}
		\mathcal{L}_{I}^j
		=
		\mathbb{E}_{o_i^{t+1}, o_{i}^{t}}\left[CE\left[p\left(z_j \mid o_{i}^{t}\right) \| q_{\phi}\left(z_j \mid o_i^{t+1}, o_{i}^{t}\right)-H\left(z_j \mid o_{i}^{t}\right)\right] \right.
		\\= \mathbb{E}_{o_i^{t+1}, o_{i}^{t}}\left[D_{\mathrm{KL}}\left[p\left(z_j \mid o_{i}^{t}\right) \| q_{\phi}\left(z_j \mid o_{i}^{t+1}, o_{i}^{t}\right)\right]\right].
	\end{array}
\end{equation}
Then the lower bound can be estimated by optimizing this loss of each team. The proof is finished. \qed

Furthermore, the joint intentions also should connect with the observations of other agents in the team. For example, joint intention of forcing fire requires all agents in the team to approach the target enemy, which means the distance between an agent and the enemy should decrease in its observation. We utilize another mutual information $I(o_k^{t+1};z_j)$ to estimate this connection. 
Additionally, we find that this mutual information is crucial for the cooperation in homogeneous settings. 
As all agents are potential commanders, it is important that all agents in the same team have a similar understanding of the environment to produce the same joint intention. The corresponding loss is used as an auxiliary loss to accelerate the training process, which is
\begin{equation}
	\mathcal{L}_{A}^j= \frac{1}{K_j}\sum_{k=0}^{K_j} \mathbb{E}_{o_{i}^{t},o_{k}^{t+1}}\left[D_{\mathrm{KL}}\left[p\left(z_j \mid o_{i}^{t}\right) \| q_{\phi}\left(z_j \mid o_{k}^{t+1}, o_{i}^{t}\right)\right]\right].
\end{equation}

However, these optimized objectives so far do not guarantee diversity of the joint intentions, which means each joint intention is expected to describe one specific change of the local state. To address this, we propose another loss to minimize the mutual information between $o_{i}^{t+1}$ and $z_j^-$, where $z_j^-$ means the unselected $z$ in the latent space. Thus, the loss is 
\begin{equation}
	\mathcal{L}_{D}^j= -\mathbb{E}_{o_{i}^{t},o_{i}^{t+1}}\left[D_{\mathrm{KL}}\left[p\left(z_j^- \mid o_{i}^{t}\right) \| q_{\phi}\left(z_j^- \mid o_{i}^{t+1}, o_{i}^{t}\right)\right]\right].
\end{equation}

Finally, both joint intention and behavior policies are built on the structure of QMIX class method. 
Behavior policy uses the typical QMIX structure while joint intention policy uses VDN network to address the variable number of teams.
And to compensate for the loss of expressiveness of VDN and overcome the non-stationarity of hierarchical framework, we propose a novel mixing method for joint intention policy which will be discussed in the following section. 
The TD-error loss of joint intention and behavior policies are
\begin{equation}
	\begin{array}{c}
		\mathcal{L}_{TD}^{l}=\mathbb{E}_{\pi_{l}}[Q_{tot}^{l}(s_{t}, \mathbf{z_t}, \mathbf{u}_{t})-y_t^l]^2
		\\
		y_t^l=r_t+\gamma \max _{\mathbf{u}_{t+1}} Q_{tot}^{l}\left(s_{t+1}, \mathbf{z_{t+1}}, \mathbf{u}_{t+1}\right)
		\\
		\mathcal{L}_{TD}^{h}=\mathbb{E}_{\pi_{h}}[Q_{tot}^{h}(s_{t}, \mathbf{z}_{t})-y_t^h]^2
		\\
		y_t^h=r_t+\gamma \max _{\mathbf{z}_{t+1}} Q_{tot}^{h}\left(s_{t+1}, \mathbf{z}_{t+1}\right).
	\end{array}
\end{equation}

Combining all these objective we introduced, the final learning objective of our method is:
\begin{equation}
	\begin{array}{c}
		\mathcal{L}_{TD}^h(\theta)+\mathcal{L}_{TD}^l(\mu)+\sum_{j=0}^{j=M}(\mathcal{L}_{I}^j(\mu,\phi)+\lambda_1 \mathcal{L}_{A}^j(\mu,\phi) +\lambda_2\mathcal{L}_{D}^j(\mu,\phi)).
	\end{array}
\end{equation}
where $M$ is the number of teams, $\lambda_1, \lambda_2$ are positive multipliers to control the rate of optimization and $\theta, \mu, \phi$ are parameters of networks.

\subsection{Weighted Value Decomposition}
However, there is another problem that we can not directly use the mixing network of QMIX for high level joint intention policy, because the mixing network of QMIX have a fixed dimension of inputs and the numbers of teams are dynamically changing. 
Fortunately, we can find an alternative mixing method to address the value decomposition problem. Firstly, we consider reward decomposition at the team level. We have $Q_j$ stands for the joint intention policy's value function of each team, which is the value function of team commander's joint intention policy under fixed behavior policies $\pi_l^*$ of all agents in the team. If we have $r_{tot}(s,\mathbf{z})= \sum_{j=0}^{j=M} r_j(o_i^t,z_j)$, where $r_j$ means the reward decomposition of team $j$ based on local state and joint intention of the team. 
\begin{equation}
	\begin{array}{c}
		Q_{tot}^h(s,\mathbf{z}) =\mathbb{E}\left[\sum_{t=0}^{\infty} \gamma^{t} r\left({s}, \mathbf{z}\right) \mid \pi_h,\pi_l^* \right] 
		= \mathbb{E}\left[\sum_{t=0}^{\infty} \gamma^{t} \sum_{j=0}^M(r_j(o_i^t,z_j)) \mid \pi_h,\pi_l^*\right]                          \\=\sum_{j=0}^M Q_{j}(s,\mathbf{z})
		\approx \sum_{j=0}^M Q_{j}(o_i^t,z_j).
	\end{array}
\end{equation}
VDN indicates the approximation is because one team's expected future return could be expected to depend more strongly on observations and actions due to the team itself than those due to other teams. However, such an approximation is oversimplified. Unlike QMIX proposes a hyber-network \cite{ha2016hypernetworks} conditioning on the global state as a more expressive factorization, our method utilizes the mutual information to estimate the importance of each value function and mixing them. Specifically, our mixing method is 
\begin{equation}
	\begin{array}{c}
		Q_{tot}^h(s,\mathbf{z})=\sum_{j=0}^M Q_{j}(s,\mathbf{z})
		\approx \sum_{j=0}^M \alpha_j Q_{j}(o_i^t,z_j) \\
		\alpha_j = \frac{K_j \cdot I(o_i^{t+1} ; z_j \mid o_i^{t})}{\sum_{j=0}^{j=K_j}(I(o_i^{t+1} ; z_j \mid o_i^{t}))}.
	\end{array}
\end{equation}
where $\alpha_j$ is the weights of each factor. We find that the mixing form is equivalent to assigning different gradients to different value function learning samples.

\textbf{Proof.}

For the weighted value decomposition, the gradients are
\begin{equation}
	\begin{array}{c}
		\mathcal{L}_{TD}^{h}(s_{t}, \mathbf{z}_{t})
		 = \mathbb{E}_{\pi_{h}}[Q_{tot}^{h}(s_{t}, \mathbf{z}_{t})-(r_t+\gamma \max _{\mathbf{z}_{t+1}} Q_{tot}^{h}\left(s_{t+1}, \mathbf{z}_{t+1}\right))]^2
		\\
		=[\sum_{j=0}^M \alpha_j Q_{j}(o_i^t,z_{t})-(r_t+\gamma \sum_{j=0}^M \alpha_j Q_{j}^{T}(o^{t+1}_i,z^{*}_j))]^2.
	\end{array}
\end{equation}
where $z^{*}_j$ means the optimal action and $Q_{j}^{T}$ is the target network. Notably, we use the same $\alpha_j$ in the target Q-value to stabilize the training process. We can transfer it into another formulation
\begin{equation}
	\begin{array}{c}
		\mathcal{L}_{TD}^{h}(s_{t}, \mathbf{z}_{t})= [\sum_{j=0}^M (\alpha_j Q_{j}(o_i^t, z_{t})-\frac{r_t}{M}-\gamma \alpha_j Q_{j}^{T}(o^{t+1}_i,z^{*}_j))]^2.
	\end{array}
\end{equation}
If we have
\begin{equation}
	\begin{array}{c}
		y_{TD} = \sum_{j=0}^M (\alpha_j Q_{j}(o_i^t, z_{t})-\frac{r_t}{M}-\gamma \alpha_j Q_{j}^{T}(o^{t+1}_i,z^{*}_j)).
	\end{array}
\end{equation}
Then
\begin{equation}
	\begin{array}{c}
		\frac{\partial \mathcal{L}_{TD}^{h}(s_{t}, \mathbf{z}_{t})}{\partial \theta}= 2 \cdot y_{TD} \cdot \sum_{j=0}^M \alpha_j \frac{\partial Q_{j}}{\partial \theta}.
	\end{array}
\end{equation}
We find that $\alpha_j$ plays a similar role as the learning rate and can control the value of the gradient. Therefore, we prove that our method equals to decrease the learning rate when the samples have a lower mutual information. The proof is finished. \qed

Therefore, the insight behind this mixing is that the team whose mutual information is higher is supposed to produce more meaningful joint actions, and the corresponding posterior may have a good estimation. Thus, our method assigns a higher gradients to learn the samples of these teams. 
Moreover, since the non-stationarity is caused by samples generated by not well-trained low level policy, such samples have lower mutual information thus have a lower gradient in our method, which will reduce the impact of non-stationary samples on high level policy. This shows that our mixing method can also reduce the non-stationarity of hierarchical framework. 

\section{Experiments}
\subsection{Environments and Experimental Settings}
For the experiments, we conduct experiments in both monotonic environments and non-monotonic environments to test all methods. Specifically, we adopt a challenging set of cooperative StarCraft II maps from the SMAC benchmark as monotonic environments \cite{samvelyan2019starcraft} and several scenarios from grid-world platform MAgent as non-monotonic environments \cite{zheng2018magent}.
In SMAC, all maps have classified as Easy, Hard and Super Hard. We use two hard scenarios \textit{3s\_vs\_4z} and \textit{5m\_vs\_6m} as well as two super hard scenarios \textit{corridor} and \textit{6h\_vs\_8z} for experiments. The optimal cooperative actions do not depend on each other's actions in SMAC \cite{gupta2021uneven}. 
In MAgent, we choose two different scenarios \textit{pursuit} and \textit{tiger}. The tasks in both scenarios require all agents to take the cooperative joint actions to catch the prey, otherwise the agents receive a penalty for taking the catching action alone. Therefore these scenarios are non-monotonic environments. 
\begin{table}[t]
	\centering
	\caption{Settings of MAgent scenarios.}
	\scalebox{0.95}{
		\begin{tabular}{|c|c|c|c|c|c|c|}
			\hline              & Pursuit        & Pursuit hard     & Tiger          \\
			\hline Agent number & 6              & 6                & 6              \\
			\hline Enemy number & 4              & 6                & 24             \\
			\hline Map size     & 60 $\times$ 60 & 100 $\times$ 100 & 40 $\times$ 40 \\
			\hline Wall number  & 60             & 300              & 60             \\
			\hline
		\end{tabular}
	}
	
	\label{table2}
\end{table}
Additionally, preys in \textit{tiger} can recover HP point at each time step, so the agents should learn to let the preys go instead of killing it immediately to get higher return. There are the detailed settings of these scenarios, as shown in Table \ref{table2}. 
The global state of MAgent is a mini map ($10 \times 10$) of the global information. The opponent's policies used in experiments are randomly escaping policy in \textit{pursuit} and \textit{tiger}. 
We use the performance of evaluation episodes with greedy action selection as the final performance. The performance is evaluated by win rate in SMAC and by return reward divided by the number of the agents in MAgent. All experiments are carried out with five random seeds. 

In the experiments, we compare our method with G2ANET, QMIX, MAVEN and QTRAN. G2ANET is a communication method while others are decentralized methods. To ensure the comparison is fair, all methods use the same basic hyperparameters and network structures with similar parameters. The network of all compared methods and our behavior policy uses the same LSTM network, consisting of a recurrent layer comprised of a GRU with a 64-dimensional hidden state, with one fully-connected layer before and two after. The network of our intention policy uses two fully-connected layers with 64-dimensional hidden state, with one fully-connected 32-dimensional layer after. The intention latent space's size is 16 in all experiments. Specially, G2ANET use hard attention which is a recurrent layer comprised of a GRU with a 64-dimensional hidden state, with one fully-connected layer after.

As for hyperparameters, we set the discount factor as 0.99 and use the RMSprop optimizer with a learning rate of 5e-4. The $\epsilon$-greedy is used for exploration with $\epsilon$ annealed linearly from 1.0 to 0.05 in 70k steps.
The batch size is 4 and updating the target every 200 episodes. The length of each episode in MAgent is limited to 350 steps while in SMAC is unlimited. The $\lambda_1, \lambda_2$ that are used to control the optimization ratio of different objective are both 1.0. All experiments are carried out on the same computer, equipped with an Intel i7-7700K, 32GB RAM and an NVIDIA GTX1080Ti. The system is Ubuntu 18.04 and the framework is PyTorch. 

\begin{figure*}[t]
	\centering
	\includegraphics[width=1.0\linewidth]{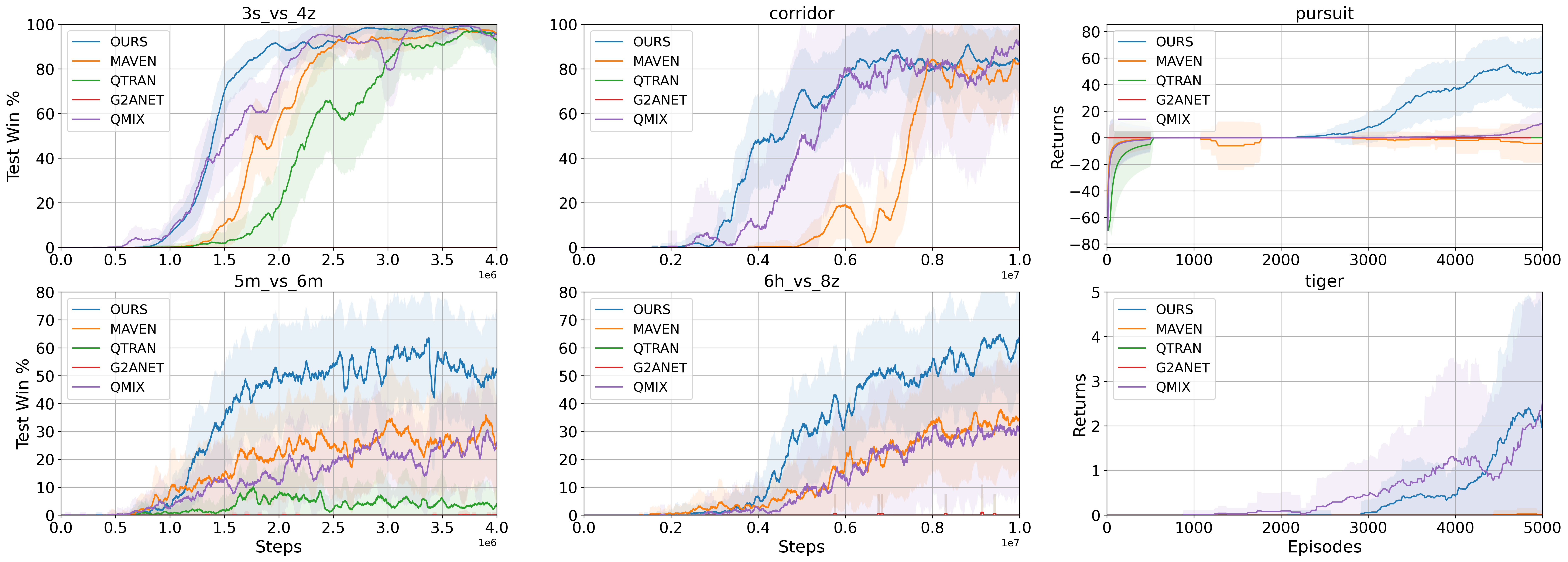}
	\caption{The results of all methods' performance. Top: The performance results in \textit{3s\_vs\_4z}, \textit{corridor} and \textit{pursuit}. Bottom: The results in \textit{5m\_vs\_6m}, \textit{6h\_vs\_8z} and \textit{tiger}.}
	\label{rl}
\end{figure*}

\subsection{Performance}
We show the results of six different scenarios from SMAC and MAgent in Fig. \ref{rl}. 
The results show that our method performs substantially better than all compared methods in all scenarios. In the monotonic environment, we notice that QMIX does not suffer from the monotonic restriction and has a relatively better performance. However, it still fails in the scenarios where the tasks require agents to act simultaneously such as focus fire in \textit{6h\_vs\_8z}. In contrast, our method can autonomously learn joint intentions to capture the interactions among agents and take the corresponding optimal cooperative actions. 
Moreover, we notice that our method has a more significant advantage in non-monotonic environments, like \textit{pursuit} and \textit{tiger} than in monotonic environments like SMAC maps. 
The reason is that tasks in non-monotonic environments require agents to take optimal actions based on each others' actions. All compared methods can not consider others' actions, so they learn a lazy policy that never attacks to avoid penalty. However, our method uses the joint intention to represent the common goal which makes all agents know that other agents will take the cooperative actions corresponding to the common goal. Therefore, cooperation can be achieved. This result indicates that our method indeed addresses the relative overgeneralization problem by following the joint intentions.
However, we find that G2ANET performs less satisfactorily in most scenarios. We believe this is because G2ANET has more hyper-parameters and therefore may lack generalization capabilities.

\begin{figure*}[t]
	\centering
	\includegraphics[width=1.0\linewidth]{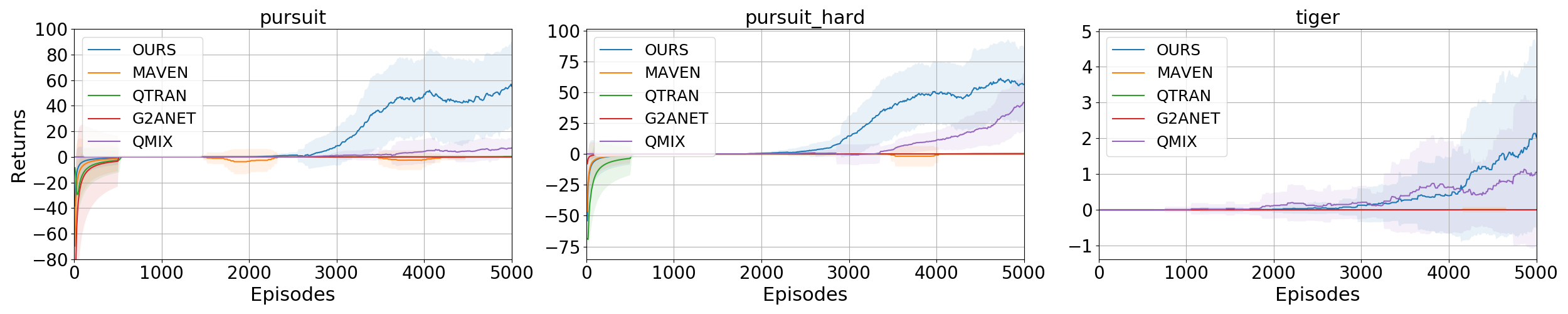}
	\caption{The results of all methods' performance in ad hoc team play settings. Left: The performance results in \textit{pursuit}. Middle: The performance results in \textit{pursuit hard}. Right: The performance results in \textit{tiger}.}
	\label{ad}
\end{figure*}

\subsection{Policy Transfer in Ad Hoc Team Play Settings}
Since the learned joint intentions are independent of the dynamically changing number of agents in the team, our method is expected to learn a flexible policy to handle cooperation with different numbers of agents. Therefore, we evaluate all methods in ad hoc team play settings. 
Specifically, we use scenarios that have the same tasks as the training environment but a different number of agents to evaluate all methods. 
We change the number by using a random number in the range of plus 2 to minus 2, which is 4-8 in \textit{pursuit}, \textit{pursuit hard} and \textit{tiger}. The rest of implementations of all ad hoc team play scenario settings are the same as shown in Table \ref{table2}. 
The result in Fig. \ref{ad} shows that our method has no significant performance degradation in ad hoc team play settings. However, other methods suffer from the various number of agents and the performance is not comparable to in the original settings. This result indicates that the joint intentions of our method can describe the interactions among various agents, enabling the learned policy to adapt to ad hoc team play scenarios. 


\subsection{Interpretability of Joint Intentions}
\begin{figure*}[t]
	\centering
	\includegraphics[width=0.8\linewidth]{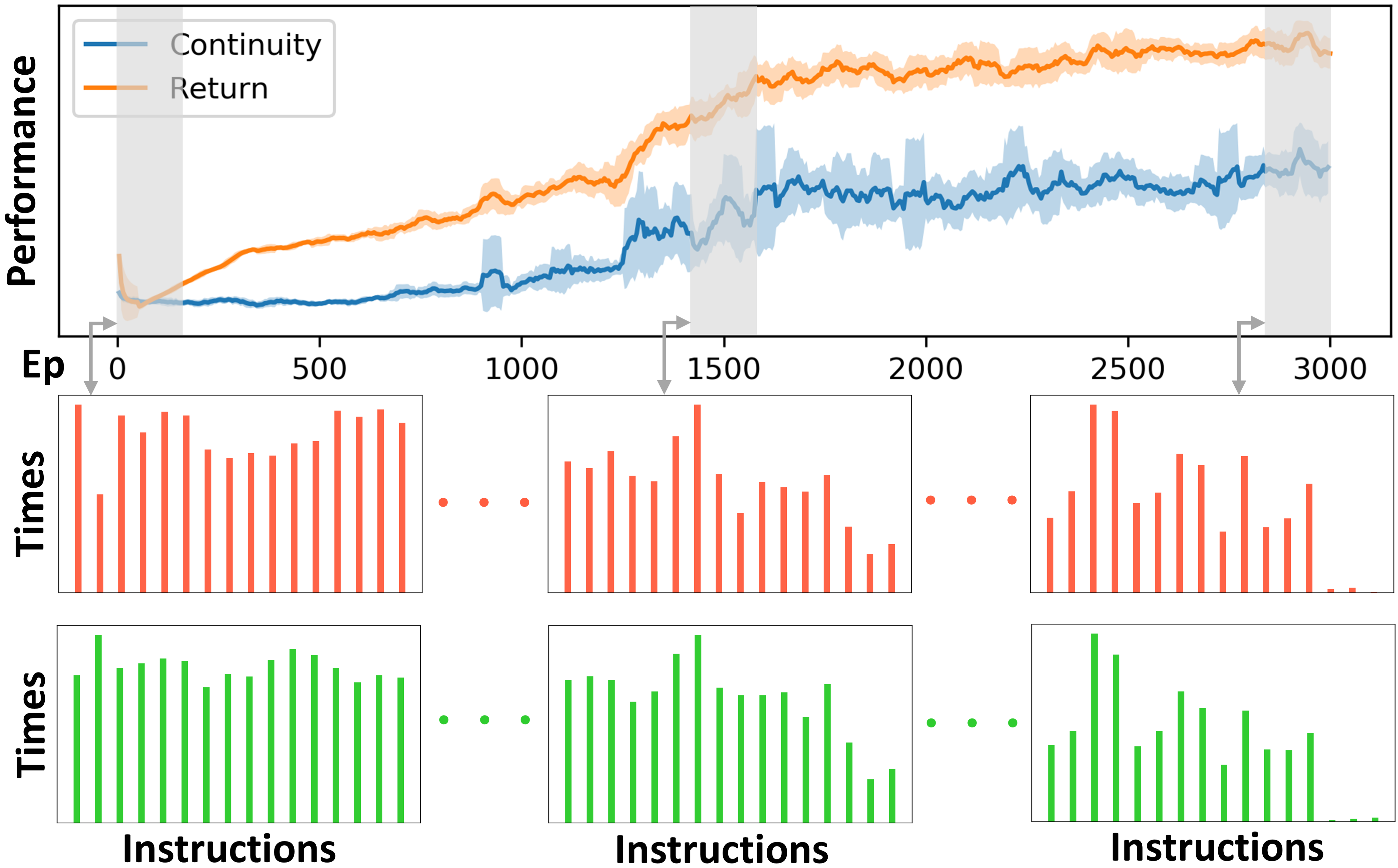}
	\caption{Top: The result of policy's performance and the continuity of joint intentions during the training process. Middle (Red): The result of the number of times each joint intention is selected in an episode from initial (left) to final (right) of the training process. Bottom (Green): The result of the number of times each joint intention is selected based on each agent's observation instead of the commander's observation in an episode from initial (left) to final (right) of the training process. All result has been normalized for better representation.}
	\label{ins}
\end{figure*}
\begin{figure}[t]
	\centering
	\includegraphics[width=0.8\linewidth]{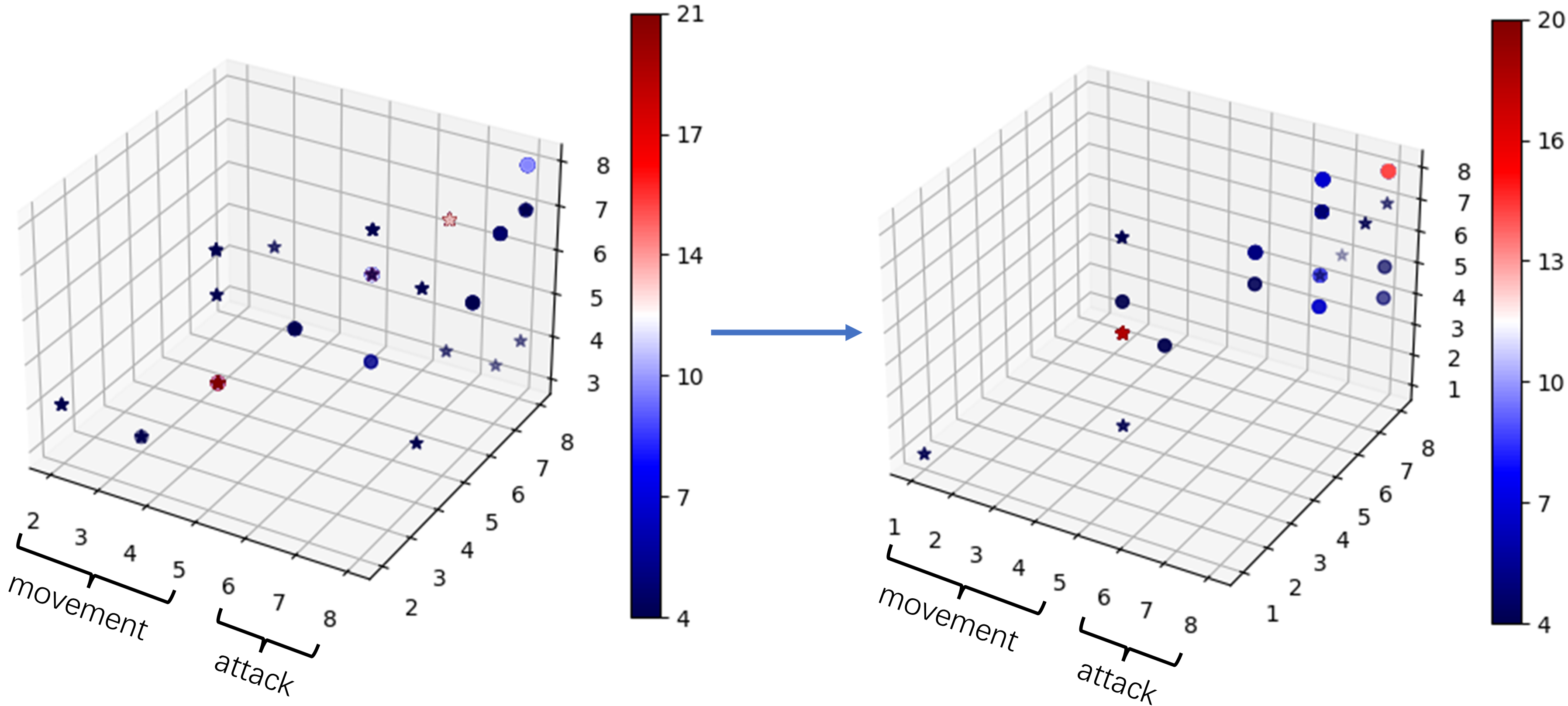}
	\caption{Representation of learned joint intentions from initial (Left) to final (Right) of the training process. Different marks represent different joint intentions and the color bar represents the number of occurrences of each joint intention in 20 episodes (closer to red means larger number). All axes are the agents' actions, and we show the actions related to movement and attack in the figure.}
	\label{uz}
\end{figure}
\subsubsection{Emergence of Joint Intentions}
We show the training process of \textit{pursuit} along with the changes of distributions of joint intentions in Fig. \ref{ins}. We analyze three important distributions to show interpretability of our method. The first one is the number of times each joint intention has been selected in an episode. The result shows that in the early stage of training, the selection of joint intentions is basically random. As the training progresses, the joint intentions’ meaning becomes clearer so that the choices become differentiated. This result indicates that the joint intentions learned by our method are indeed meaningful. 

The second distribution is the number of times each joint intention has been selected based on each agent's observation instead of the commander's observation in an episode. The result shows that this distribution gets more similar to the actual joint intentions' distribution as training progresses. 
Such results show that the agent's knowledge of the task tends to converge during training. Since the task is fixed, the part of agents that are not selected as commanders but can observe enough information to recognize the surrounding state will have the same intentions as the commanders to solve the task. This also indicates that the learned intentions are meaningful, otherwise agents would not have consistent intentions for the same task. 

The last distribution is the continuity of joint intentions. We use the number of consecutive executions of the same joint intention by the agents as the result. The result shows that agents learn to choose and follow a single joint intention for more steps as training progresses. This continuity feature indicates that the learned joint intentions are meaningful and can represent some long-term goals while agents can take corresponding responses to achieve these goals. Furthermore, we find that the continuity curve converges as the performance curve converges. This result indicates that the learning process of joint intentions is related to performance improvement of the policy, which proves that joint intentions can promote performance improvement.


\subsubsection{Representation of Learned Joint Intentions}

To explain the interpretability of our method, we also present how the behaviors of agents in \textit{3m} of SMAC scenarios are related to the learned joint intentions representations in Fig. \ref{uz}. We show the joint actions of all three agents and the corresponding choices of joint intentions. At the beginning of training, the joint actions are irrelevant to the joint intentions. As the training progresses, the joint intentions become more meaningful and the final joint intentions clearly represent two main behavior modes. 
The results show the circle marks mainly emerge in the area where all agents are taking attack actions. This means such joint intention represents the behavior mode that the agents choose to attack the target together. Similarly, the star marks are mainly in the movement area and represent the joint intention that the agents choose to move together. 
\begin{table}
	\small
	\centering
	\caption{Ablation experiments on evaluating without joint intentions.}
	\begin{tabular}{|c|c|c|c|c|}
		\hline & Our method & Our method w/o joint intentions \\
		\hline 6h\_vs\_8z & 100\%  & 0\%\\
		\hline corridor & 100\%  &0\% \\
		\hline 5m\_vs\_6m & 100\%  &0\% \\
		\hline pursuit & 100\%  &56\% \\
		\hline
	\end{tabular}
	\label{t1}
\end{table}
\begin{figure}[t]
	\centering
	\includegraphics[width=0.9\linewidth]{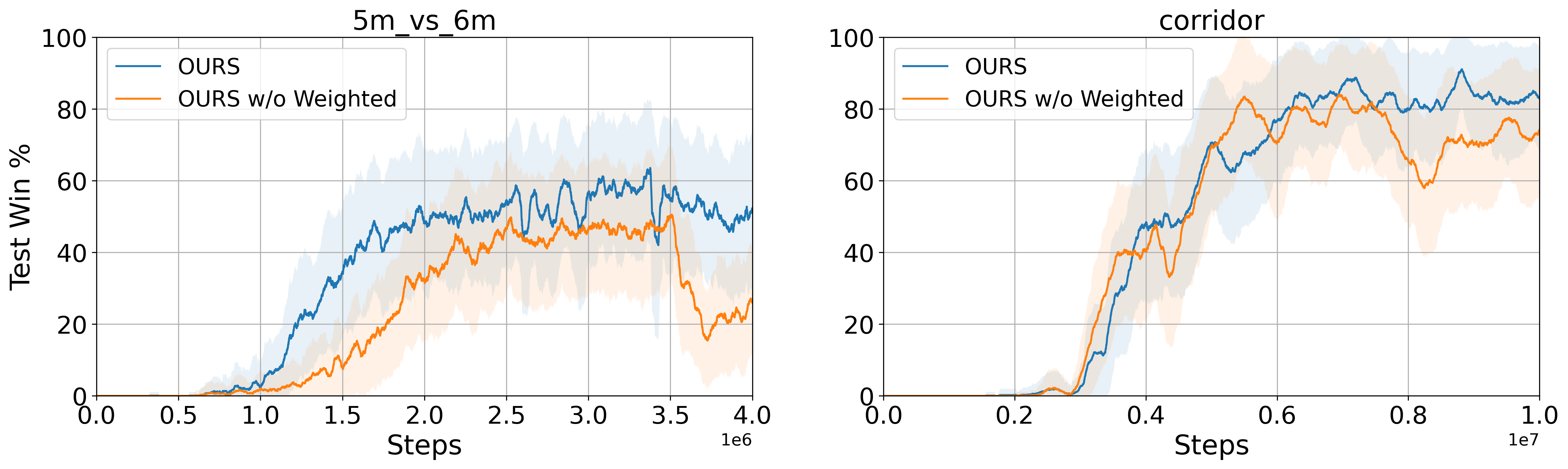}
	\caption{Ablation experiments on training without weighted value decomposition.}
	\label{alt}
\end{figure}

\subsection{Ablations}
\subsubsection{Ablation of Joint Intentions}
\label{dvd}
To evaluate how the performance of our policy relies on the joint intentions, we evaluate our method with empty joint intentions. We normalize the performance of all results with the performance of our method. The results are shown in Table \ref{t1}. Every experiment is carried out with 100 episodes. 
The result indicates that cooperation requires following meaningful joint intentions to achieve as the performance that executed without learned joint intentions is close to that of the random policy.

\subsubsection{Ablation of Weighted Value Decomposition}
We perform several ablations on the \textit{5m\_vs\_6m} and \textit{corridor} scenarios. We consider training without the weighted value decomposition. The ablations results are shown in Fig. \ref{alt}. The experiment shows that the weighted value decomposition improves the stationarity as the policy trained without it has a significant decrease in the final performance.

\section{Conclusion}
In this work, we propose a novel multi-agent reinforcement learning method based on a hierarchical framework and learns unsupervised joint intentions to tackle the relative overgeneralization problem. 
We use a proposed team partition method which can minimize the optimality gap of rewards to share joint intentions among each team. Then, we use a hierarchical framework to learn a high level joint intention policy and a low level behavior policy. 
The joint intentions can be unsupervised autonomously learned in a latent space and behavior policy can learn to take optimal cooperative actions corresponding to joint intentions. 
Moreover, we propose a weighted value decomposition using mutual information to overcome the non-stationarity in hierarchical reinforcement learning. We also illustrate that our method can adapt to ad hoc team play situations. 
The experiments show that our method achieves significant performance improvements by learning meaningful joint intentions in all domains and we illustrate that joint intention learning is related to policy's performance improvement which showcases the efficiency of the learned joint intentions. 
\bibliographystyle{plain}
\bibliography{reference}
\end{document}